\documentclass[
reprint,
bibnotes,
amsmath,amssymb,
aps,
prd, tikz
]{revtex4-2}

\usepackage{graphicx}
\usepackage{dcolumn}
\usepackage{enumitem}
\usepackage{bm}
\usepackage{float}
\usepackage[many]{tcolorbox}
\usepackage{shuffle}
\usepackage{environ}
\usepackage{physics}
\usepackage{tabularx}
\usepackage{diagbox}
\usepackage{subcaption}
\usepackage{multirow}
\usepackage{hyperref}

\newcommand{\cym}{C^{\mathrm{YM}}}

\usepackage{CJK}

\usepackage{tikz}
\usetikzlibrary{mindmap,trees,positioning,quotes,shapes.geometric,arrows,decorations.pathmorphing,decorations.markings,patterns,calc}

\begin{document}

\begin{CJK*}{UTF8}{}
\CJKfamily{gbsn}

\title{
Loop-Level Double Copy Relations from Forward Limits}

\author{Qu Cao(曹趣)$^{1,2}$}
\email{qucao@zju.edu.cn}
\author{Song He(何颂)$^{2,3}$}
\email{songhe@itp.ac.cn}
\author{Yong Zhang(张勇)$^{4}$}
\email{zhangyong1@nbu.edu.cn}
\author{Fan Zhu(朱凡)$^{5}$}
\email{zhufan25@gscaep.ac.cn}

\affiliation{
$^{1}$Zhejiang Institute of Modern Physics, School of Physics and Joint Center for Quanta-to-Cosmos Physics, Zhejiang University, Hangzhou, Zhejiang 310027, China \\
$^{2}$Institute of Theoretical Physics, Chinese Academy of Sciences, Beijing 100190, China \\
$^{3}$School of Fundamental Physics and Mathematical Sciences, Hangzhou Institute for Advanced Study and ICTP-AP, UCAS, Hangzhou 310024, China\\
$^{4}$
 Institute of Fundamental Physics and Quantum Technology\\ \& School of Physical Science and Technology, Ningbo University, Ningbo, Zhejiang 315211, China 
\\
$^5$Graduate School of China Academy of Engineering Physics, No. 10 Xibeiwang East Road, Haidian District, Beijing,
100193, P.R.China}

\begin{abstract}
We study double copy relations for loop integrands in gauge theories and gravity based on their constructions from single cuts, which are in turn obtained from forward limits of lower-loop cases. While such a construction from forward limits has been realized for loop integrands in gauge theories, we demonstrate its extension to gravity by reconstructing one-loop gravity integrands from forward limits of trees. Under mild symmetry assumptions on tree-level kinematic numerators (and their forward limits), our method directly leads to double copy relations for one-loop integrands: these include the field-theoretic Kawai-Lewellen-Tye (KLT) relations, whose kernel is  the inverse of a matrix with rank $(n{-}1)!$ formed by those in bi-adjoint $\phi^3$ theory, and the Bern-Carrasco-Johansson (BCJ) double copy relations with crossing-symmetric kinematic numerators (we provide local and crossing-symmetric Yang-Mills BCJ numerators for $n=3,4,5$ explicitly). By exploiting the ``universal expansion" for one-loop integrands in generic gauge theories, we also obtain an analogous expansion for gravity (including supergravity theories).

\end{abstract}
\maketitle
\end{CJK*}

\noindent {\bf Introduction.}

Recent years have witnessed enormous progress on the study of scattering amplitudes in gauge theories and gravity, including numerous new structures and especially new relations such as color-kinematics duality and the double copy construction~\cite{Bern:2008qj,Bern:2010ue} ({\it c.f.} \cite{Bern:2019prr, Bern:2022wqg,Adamo:2022dcm,Bern:2023zkg} for recent reviews and references therein). ~
At tree level, double copy relations have been proven rigorously and they are equivalent to the manifestly gauge-invariant, field-theoretic Kawai-Lewellen-Tye (KLT) relations, which can be obtained as the $\alpha' \to 0$ limit of the original KLT relations between open- and closed-string amplitudes~\cite{Kawai:1985xq}. It expresses the $n$-point gravity tree amplitude as a bilinear of color-ordered gauge-theory amplitudes in a minimal
 $(n{-}3)!$ basis, with the field-theoretic KLT kernel~\cite{Bern:1998sv, Stieberger:2009hq, Bjerrum-Bohr:2009ulz}
given by the inverse of the matrix (with rank $(n{-}3)!$) formed by the so-called bi-adjoint $\phi^3$ amplitudes~\cite{Cachazo:2013gna, Cachazo:2013hca, Cachazo:2013iea, Cachazo:2014xea}. 

Despite enormous recent progress ({\it c.f.} \cite{Bern:2024vqs}), it is fair to say that double copy relations still remain conjectural at loop level in general. In an attempt at one-loop level in~\cite{He:2016mzd, He:2017spx}, KLT/BCJ double copy relations were derived for loop integrands with linear-in-$\ell$ propagators, which are essentially inherited from tree-level relations by taking  ``forward limits" of trees with a pair of legs in higher dimensions~\cite{Baadsgaard:2015twa,He:2015yua,Cachazo:2015aol}; such integrands are also naturally obtained from CHY formulas and underlying ambitwistor string theory~\cite{Geyer:2015bja,Geyer:2015jch,Geyer:2017ela} and can be extended to supersymmetric gauge theories and supergravity by including fermions/scalars in the loop~\cite{Edison:2020uzf}. In fact, by exploiting these results one can bootstrap BCJ numerators up to $n=7$ for maximally supersymmetric cases (and up to $n=5$ for half-maximal cases)~\cite{Edison:2022jln}. It remains an important open question if KLT/BCJ double copy relations can be obtained for one loop ``physical" integrands with standard Feynman propagators (quadratic in $\ell$), and the main goal of this letter is to give an affirmative answer to this question. 

In a separate line of research, a new method inspired by ``surfaceology"~\cite{Arkani-Hamed:2023lbd,Arkani-Hamed:2023mvg,Arkani-Hamed:2024vna,Arkani-Hamed:2024pzc} for constructing all-loop integrands of colored particles including Yang-Mills theory has been proposed based on forward limits~\cite{Arkani-Hamed:2023swr,Arkani-Hamed:2023jry,Arkani-Hamed:2024nhp,Arkani-Hamed:2024tzl}: the most remarkable feature is that the notorious issue of divergences associated with such forward limits (in non-supersymmetric theories) are canonically regulated by the ``surface kinematics" in~\cite{Arkani-Hamed:2024tzl}, which provides a canonical notion of loop integrands recursively constructed from their single cuts. In particular, as we will review shortly, it is straightforward to reconstruct the one-loop $n$-gluon YM integrand from residues which are single cuts given by forward limits of $(n{+}2)$-gluon tree amplitudes: here it is crucial that all spurious poles associated with propagators linear in $\ell$ for individual residues nicely cancel in the one-loop integrand, thus we can obtain all-multiplicity formulae for one-loop YM integrand~\cite{Cao:2024olg}.

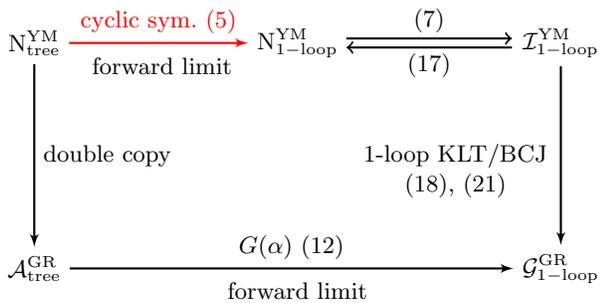
\begin{figure}[htbp]
\tikzstyle{line} = [draw, thick, -latex']
\begin{tikzpicture}[node distance = 3cm, auto,decoration={coil}]
    \node at (-1.5,0)  (nt) {${\rm N}^{\rm YM}_{\rm tree}$
    };
    \node at (2,0) (nl) { ${\rm N}^{\rm YM}_{\rm 1-loop}$};
      \node at (5.5,0) (iym) {${\cal I}^{\rm YM}_{\rm 1-loop}$
    };
    
     \node at (-1.5,-3) (agr) {${\cal A}^{\rm GR}_{\rm tree}$
    };
      \node at (5.5,-3) (ggr) {${\cal G}_{\rm 1-loop}^{\rm GR}$
    };

    \path [line,red] (nt) -- node {cyclic sym.~\eqref{cyclic-cons}} (nl);
    \node at (0.2,-0.3) {forward limit};

    \draw[->, thick] (2.65,0.06)--(4.85,0.06);
    \draw[<-, thick] (2.65,-0.06)--(4.85,-0.06); 
    \node at (3.75,0.3) {\eqref{eq_Iexpand}};
    \node at (3.75,-0.3) {\eqref{eq_Nexpand}};
      
    \path [line] (nt) -- node{double copy} (agr);
       
    \path [line] (iym) --  (ggr);
    \node at (4.1,-1.5) {1-loop KLT/BCJ};
    \node at (4.1,-1.9) {\eqref{eq:one-loop KLT},~\eqref{eq:NNM}};
       
    \path [line] (agr) --node{$G(\alpha)$\,\,\eqref{condition1}} (ggr);
    \node at (2,-3.3) {forward limit};
    
\end{tikzpicture}

\captionsetup{justification=raggedright,singlelinecheck=false}
      \caption{Starting from (kinematic numerators of) YM tree, the single-cut reconstruction based on forward limits produces one-loop integrands and BCJ numerators; similarly from gravity tree, one constructs the one-loop gravity integrands, which in turn can be obtained from one-loop KLT/BCJ double copy. }
    \label{fig_mindmap}
\end{figure}

This is the strategy we adopt in this letter: we will see that such a reconstruction can be directly extended to gravity integrands at least at one loop: while we still lack a notion of ``canonical" loop integrands for gravity, the single cuts of one-loop amplitudes can be naturally decomposed into building blocks with different orderings, which are in turn given by the forward limit of ``ordered" parts of tree amplitudes. Nevertheless, we will show that the resulting one-loop gravity integrand (after summing over all residues and contributions from all orderings) is again free of spurious poles and has all correct single cuts by construction! Once we reconstruct both integrands based on forward limits of their tree amplitudes related by double copy, we immediately obtain one-loop KLT relations inherited from those of trees, thus providing a simple solution to this longstanding problem. Our construction also leads to one-loop BCJ numerators if we assume forward limits of tree-level numerators satisfy a symmetry, and we provide explicit YM numerators at one-loop up to five points. This forward-limit reconstruction, and its interplay with double-copy structures at one loop, is schematically summarized  in Fig.~\ref{fig_mindmap}. As an application, we also derive a novel ``universal expansion" for one-loop (super-)gravity integrands from the double-copy of the well-known expansion in gauge theories (including fermions/scalars in the loop)~\cite{Cao:2024olg}. 

\noindent{\bf Forward-Limit Reconstruction of One-Loop Integrands.}

We will construct one-loop integrands in gauge theories and gravity from their single cuts, defined as forward limits of higher-point trees~\cite{Baadsgaard:2015twa, Arkani-Hamed:2024tzl, Cao:2024olg, Cao:2025mlt}.
While this guarantees the correct cuts, it also generates spurious linear propagators. We find that imposing a simple algebraic condition---\emph{shifted cyclic symmetry}---removes all such poles.
This condition, previously implicit, is both necessary and sufficient for quadratic-propagator representations, thereby yielding a direct loop-level construction of BCJ numerators from tree-level data.
The method extends naturally to gravity.

\begin{figure}[htbp]
    \centering
    \includegraphics[width=0.75\linewidth]{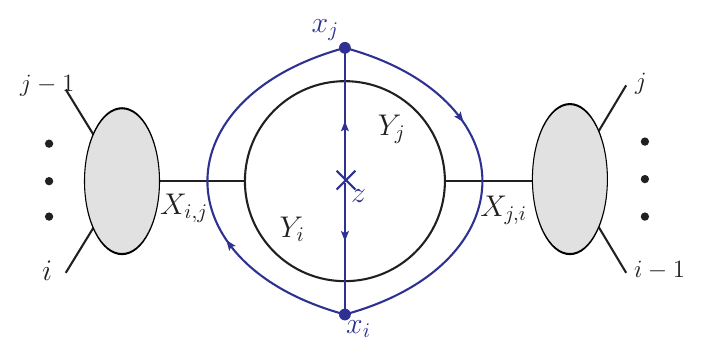}
    \captionsetup{justification=raggedright,singlelinecheck=false}
    \caption{Dual variables at one loop: the loop puncture $z$ and dual points $x_i,x_j$ define momenta $\ell_i$ along curves $z\to x_i$, giving $Y_i=\ell_i^2$. Throughout we use the in-going convention, with all external momenta flowing into the loop.}
    \label{fig_Y_var}
\end{figure}

\emph{Gauge theory.}---
For a fixed ordering of external legs $\mathbb{I}:=(1,2,\ldots,n)$, 
the poles of the Yang-Mills one-loop integrand $\tilde{\mathcal{I}}_n(\mathbb{I})$ 
include the \emph{quadratic propagators}~\footnote{We define $\ell$ to be the loop momentum preceding leg $1$, thus $\ell=\ell_1$.}
\begin{align}
    Y_i := \ell_i^2=(\ell+k_{1}+k_{2}+\cdots+k_{i{-}1})^2 ,
\end{align}
together with poles from tree sub-diagrams, e.g. 
$
X_{i,j}:=(k_i+\cdots+k_{j-1})^2,
$
as illustrated in Fig.~\ref{fig_Y_var}.  
Any one-loop integrand must contain at least one $Y_i$ in the denominator; otherwise it vanishes as a scaleless integral.

Shifting $Y_i\to Y_i(z)=Y_i-z$ and applying the residue theorem at $z=0$, the integrand can be reconstructed from
\begin{equation}
\label{residue}
   {\tilde {\mathcal{I}}}_{n}(\mathbb{I})
   =\sum_{i=1}^{n}\frac{1}{Y_i}\,
     \mathrm{Res}_{Y_i=0}\,{\tilde {\mathcal{I}}}_n\Big|_{Y_j\to Y_j-Y_i}\,,
\end{equation}
where the residues contain spurious poles of the form $Y_j-Y_i$, referred to as \emph{linearized propagators}.

\begin{figure}[htbp]
    \centering
    \includegraphics[scale=0.45]{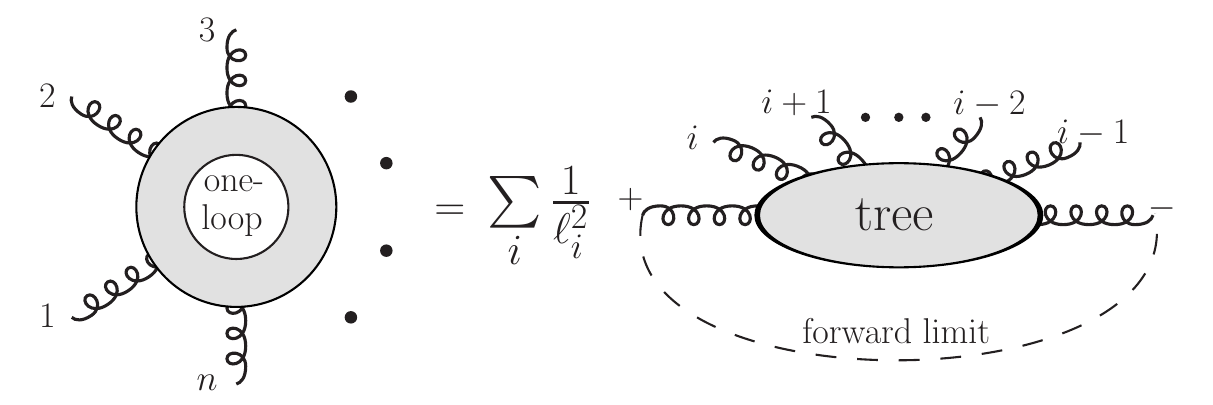}
    \captionsetup{justification=raggedright,singlelinecheck=false}
    \caption{The one-loop integrand reconstructed from forward limits of trees.}
    \label{fig_residue_reconstruction}
\end{figure}

As depicted in Fig.~\ref{fig_residue_reconstruction}, each residue is a single cut of the integrand, realized as the forward limit of an $(n{+}2)$-point tree amplitude $\mathcal{A}_{n+2}$ with a pair of momenta $K_{\pm}=\pm\ell_i$ inserted between legs $i{-}1$ and $i$. Using the surface prescription of~\cite{Arkani-Hamed:2024tzl}, the forward limit is well defined. We denote the resulting tree as $\hat{\mathcal{A}}_{n+2}(+,\cdots,-)$, where the hat indicates the state sum $\sum_{\text{states}}\epsilon_{+}^{\mu}\epsilon_{-}^{\nu}=\eta^{\mu\nu}+\ldots$. The integrand can then be written as
\begin{equation}\label{eq:I-singlecut}
   {\tilde {\mathcal{I}}}_{n}(\mathbb{I})
   = \sum_{i=1}^{n}\frac{1}{\ell_{i}^{2}}\,
     \hat{\mathcal{A}}_{n+2}(+,i,i{+}1,\ldots,i{-}1,-)\,.
\end{equation}

The forward-limit amplitude $\hat{\mathcal{A}}_{n+2}(+,i,\ldots,-)$ admits an expansion in a basis of bi-adjoint $\phi^3$ amplitudes $m_{n{+}2}(+, \pi, -\,|\, +, \rho, -)$,
\begin{align}
\label{expansion222}
\hat{\mathcal{A}}_{n+2}(+,i,\ldots,-)
   = \sum_{\rho\in S_n} {\rm N}_{\rho;\ell_i}\,
     m_{n{+}2}(+,i,\ldots,-\,|\, +,\rho,-)\,,
\end{align}
where $\rho\in S_n$ denotes the Kleiss--Kuijf basis. 
The coefficients ${\rm N}_{\rho;\ell_i}$ are forward limits of the 
$n!$ tree-level numerators~\cite{Edison:2020ehu}, 
$
{\rm N}_{\rho;\ell_i} := \hat{{\rm N}}_{n{+}2}^{\rm tree}(\ell_i,\rho,-\ell_i)\,,
$
with the hat again denoting the state sum.

Since tree-level BCJ numerators can always be chosen crossing-symmetric~\cite{Edison:2020ehu}, it follows that ${\rm N}_{\rho;\ell_i}=\rho[{\rm N}_{\mathbb{I};\ell_i}]$, where the permutation $\rho$ acts by relabeling $\{k_j,\epsilon_j\}\to\{k_{\rho_{[j]}},\epsilon_{\rho_{[j]}}\}$ while keeping $\pm\ell_i$ fixed.

Plugging the expansion \eqref{expansion222} into \eqref{eq:I-singlecut}, 
one finds that potential spurious poles of the form $Y_i-Y_j$ arise from the 
linear propagators in different $\hat{\mathcal{A}}_{n+2}(+,i,\ldots,-)$.  
A simple but crucial observation is that these poles cancel precisely when the 
kinematic numerators obey what we call a \emph{shifted cyclic symmetry},
\begin{align}\label{cyclic-cons}
 {\rm N}_{1,2,\ldots,n;\ell_1} = {\rm N}_{i,i{+}1,\ldots,i{-}1;\ell_i}\,,
\end{align}
where the loop momentum is shifted as 
$\ell_1 \to \ell_i=\ell+k_{1,\ldots,i{-}1}$.

The necessity of this condition can be seen by focusing on half-ladder 
diagrams, which only appear in 
$m(+,i,i{+}1,\ldots,i{-}1,-\,|\, +,i,i{+}1,\ldots,i{-}1,-)$ for 
$i=1,2,\ldots,n$.  
These diagrams carry linearized propagators that do not combine into quadratic 
ones unless their numerators coincide.  
Demanding such cancellation forces exactly the relations 
\eqref{cyclic-cons}, and hence shifted cyclic symmetry is unavoidable.  
Explicit examples illustrating this mechanism are given in the Appendix~\ref{app:example}.

Having established necessity, we now argue that shifted cyclic symmetry is also sufficient: once imposed, it correlates the numerators and reorganizes the forward-limit expansion into an $n!$-element basis $\mathcal{M}'$, $
{\tilde {\mathcal{I}}}_{n}(\mathbb{I})
   = \sum_{\beta\in S_{n}} {\rm N}_{\beta;\ell_1}\,\mathcal{M}'(\mathbb{I}|\beta),
$
with
\begin{equation}\label{eq:M'}
\mathcal{M}'(\mathbb{I}|\beta)
   =\sum_{i=1}^n \frac{1}{\ell_{i}^{2}}\,
      m(+,i,\ldots,i{-}1,-\,|\, +,\beta_{[i]},\ldots,-)\,.
\end{equation}
The resulting basis $\mathcal{M}'$ has been checked to involve only quadratic propagators in all explicit examples, a consequence of shifted cyclic symmetry removing the linearized poles.
Although $\mathcal{M}'$ formally consists of $n!$ elements, many are redundant: for fixed $\beta$, the coefficients ${\rm N}_{\beta;\ell_i}$ for different $i$ are related by loop-momentum shifts. Taking these shifts into account, the contributions can be reorganized into a minimal $(n{-}1)!$ basis $\mathcal{M}$, defined by collecting the shifted versions of $\mathcal{M}'$.
In this way the gauge-theory integrand takes the form
\begin{equation}
 \mathcal{I}_{n}(1,\ldots,n) =  \sum_{\gamma\in S_{n{-}1}} {\rm N}_{1,\gamma;\ell_1} \, \mathcal{M}(\mathbb{I}|1,\gamma) \cong {\tilde {\mathcal{I}}}(1,\ldots,n),
 \label{eq_Iexpand}
\end{equation}
where $\cong$ denotes equality upon integration, i.e. up to loop-momentum shifts. The minimal basis is
\begin{equation}\label{eq:M}
\mathcal{M}(\mathbb{I}|1,\gamma)
  = \sum_{i}\,
   \mathcal{M}'(\mathbb{I}|\gamma_{[i]},\ldots,1,\ldots)\Big|_{\ell_i\to\ell_1}\,.
\end{equation}
 $\mathcal{M}(\mathbb{I}|1,\gamma)$ admits a transparent diagrammatic interpretation: it collects all one-loop cubic diagrams with quadratic propagators whose external-leg ordering is compatible with both $\mathbb{I}$ and $(1,\gamma)$, with the loop momentum routed from leg $n$ to leg $1$~\cite{Feng:2022wee,Dong:2023stt}. For example, $\mathcal{M}(1,2|1,2)=\tfrac{1}{Y_1 Y_2}$ and $\mathcal{M}(1,2,3|1,3,2)=-\tfrac{1}{Y_1Y_2X_{1,2}}-\tfrac{1}{Y_1Y_3X_{1,3}}-\tfrac{1}{Y_2Y_3X_{2,3}}$.

\emph{Extension to gravity.}---
Remarkably, the same single-cut construction extends to gravity integrands. We claim that
\begin{equation}
\label{grre}
 \tilde {\mathcal{G}}= \sum_{\alpha\in S_n} {\rm N}_{\alpha;\ell} \, \tilde{\mathcal{I}}_{n}(\alpha),
\end{equation}
where the Yang-Mills integrand $\tilde{\mathcal{I}}_{n}(\alpha)$ is obtained from \eqref{expansion222} by relabeling, $\tilde{\mathcal{I}}_{n}(\alpha)=\tilde{\mathcal{I}}_{n}(\mathbb I)\big|_{\mathbb I\to \alpha}$.

Shifted cyclic  symmetry \eqref{cyclic-cons} again reduces the sum from $S_n$ to $S_{n-1}$,
\begin{equation}\label{grint}
\mathcal{G}_n =\sum_{\alpha\in S_{n{-}1}} {\rm N}_{1,\alpha;\ell} \, \mathcal{I}_{n}(1,\alpha)\cong \frac1 n  \tilde {\mathcal{G}}_n,
\end{equation}
providing a compact representation of one-loop gravity integrands on quadratic propagators.

Here we sketch a proof of \eqref{grre}.
Consider the single cut of the one-loop gravity integrand $\tilde{\mathcal{G}}_n$.
There exists a representation of $\tilde{\mathcal{G}}_n$ whose residue at $\ell^2=0$ is precisely the forward limit of the $(n{+}2)$-point gravity tree amplitude after the state sum: $\mathop{\mathrm{Res}}_{\ell^2=0}  \tilde {\mathcal{G}}_n =\sum_{\rm states}\mathcal{A}_{n{+}2}^{\rm gr}(+,\alpha,-)$. Using the Del Duca-Dixon-Maltoni  decomposition~\cite{DelDuca:1999rs} at tree level, this becomes
\begin{equation}  \mathop{\mathrm{Res}}_{\ell^2=0} {\tilde {\mathcal{G}}}_n  =\sum_{\alpha\in S_n} {\rm N}_{\alpha;\ell} \hat{\mathcal{A}}_{n{+}2}(+,\alpha,-)\,.
\end{equation}
To uplift such residues into quadratic loop integrands, we introduce partial integrands $G_n(\alpha)$, defined to satisfy
\begin{align}
\label{condition1}
      &\sum_{\alpha\in S_n}G_{n}(\alpha) = \tilde {\mathcal{G}}_n 
      , 
      \\ 
      \label{condition2}
     & \mathrm{Res}_{\ell^2=0} G_{n}(\alpha) ={\rm N}_{\alpha;\ell} \hat{\mathcal{A}}_{n{+}2}(+,\alpha,-)\,.
\end{align} 
These objects $G_n(\alpha)$ may be thought of as ordered building blocks, which sum to the full (orderless) integrand $\tilde{\mathcal G}_n$. Although gravity has no intrinsic ordering, the label $\alpha$ provides a convenient way to organize the single-cut reconstruction.

Using crossing symmetry, it suffices to take all $G_{n}(\alpha)$ as relabelings of a single seed integrand, $G_{n}(\alpha)=G_{n}(\mathbb I)\big|_{\mathbb I\to \alpha}$. Thus the problem reduces to determining $G_{n}(\mathbb{I})$.
Adapting the residue-theorem trick used in \eqref{eq:I-singlecut}, we shift $Y_i\to Y_i-z$ and evaluate the contour at $z=0$, 
\begin{equation}
\label{residueG}
   G_{n}(\mathbb I)=\sum_{i=1}^{n} 
    \frac{1}{Y_i} \mathrm{Res}_{Y_i=0}G_{n}(\mathbb I)\Big|_{Y_j\to Y_j-Y_i}\,.
\end{equation}
Invoking condition \eqref{condition2}, each residue evaluates to
\begin{align}
&\mathrm{Res}_{Y_i=0}G_{n}(\mathbb I) (Y_j\to Y_j-Y_i)
\nonumber \\ & \qquad={\rm N}_{i \cdots i-1;\ell_i} \hat{\mathcal{A}}_{n{+}2}(+,i \cdots i-1,-).
\end{align}
By shifted cyclic symmetry \eqref{cyclic-cons}, the numerators ${\rm N}_{i\cdots i-1;\ell_i}$ reduce to a common overall factor. This yields
\begin{align}
  G_{n}(\mathbb I)
=    {\rm N}_{\mathbb I;\ell_1}  \sum_{i=1}^{n}
     \frac{\hat{\mathcal{A}}_{n+2}(+,i,\ldots,i{-}1,-)}{\ell_{i}^{2}}  = {\rm N}_{\mathbb I;\ell_1} \tilde{\mathcal{I}}_{n}(\mathbb I),
\end{align}
where in the last step we used the Yang-Mills identity \eqref{eq:I-singlecut}.
Relabeling $G_{n}(\mathbb{I})$ and imposing condition \eqref{condition1} then immediately reproduces \eqref{grre}.

Next we take Eq.~\eqref{grint} as a starting point to derive the one-loop KLT/BCJ double-copy relations.

\noindent {\bf One-loop KLT/BCJ double copy relations.}

\paragraph{KLT relations} 
Our construction yields one-loop KLT relations directly. The key step is to rewrite the numerators ${\rm N}_{1,\alpha;\ell}$ in~\eqref{eq_Iexpand} in terms of gauge-theory integrands $\mathcal{I}_n$, without requiring their explicit form. Since the matrix ${\cal M}(1,\alpha|1,\beta)$ has full rank $(n{-}1)!$, \eqref{eq_Iexpand} can be inverted, giving
\begin{equation}
    {{\rm N}}_{1,\alpha;\ell}=\sum_{\beta\in S_{n-1}}\mathcal{K}(1,\alpha|1,\beta)\mathcal{I}_n(1,\beta)\,,
    \label{eq_Nexpand}
\end{equation}
where the one-loop KLT kernel is defined as the inverse matrix, $\mathcal{K}(1,\alpha|1,\beta):=(\mathcal{M}_{n}(1,\alpha|1,\beta))^{-1}$. Substituting \eqref{eq_Nexpand} into \eqref{grint} gives the one-loop KLT relation,
\begin{equation}\label{eq:one-loop KLT}
 \mathcal{G}_n =\sum_{\alpha,\beta\in S_{n{-}1}} \mathcal{I}_{n}(1,\alpha) \mathcal{K}(1,\alpha|1,\beta)\mathcal{I}_{n}(1,\beta)\,.
\end{equation} 
Eq. \eqref{eq:one-loop KLT} is one of our main results. Although we have illustrated it using Yang-Mills theory and its double copy, we conjecture that the relation is universal and applies to any gauge theory and its corresponding gravity theory (including supersymmetric cases)~\cite{Edison:2020uzf}. An important feature is that no explicit one-loop BCJ numerators are required: for any representation of the color-ordered gauge-theory integrands, the KLT formula automatically yields a valid gravity integrand since all single cuts are guaranteed. In contrast to earlier one-loop formulations involving linear-propagator representations, our relation is built from local gauge-theory integrands expressed purely in terms of trivalent graphs with quadratic propagators.

For $n=2$, the only contribution is the bubble diagram, and the kernel reduces to a $1\times1$ matrix,
\begin{equation}
     \mathcal{G}_2= \mathcal{I}_{1,2} (\frac{1}{P_{1,2}})^{-1}\mathcal{I}_{1,2}=\frac{{\rm N}_{1,2;\ell}}{P_{1,2}}(\frac{1}{P_{1,2}})^{-1}\frac{{\rm N}_{1,2;\ell}}{P_{1,2}}=\frac{{\rm N}_{1,2;\ell}^2}{P_{1,2}}\,,
\end{equation}
where $P_{1,2}=\ell_1^2\ell_2^2$ denotes the product of the two quadratic propagators.

\begin{figure}[htbp]
    \centering
    \includegraphics[width=0.99\linewidth]{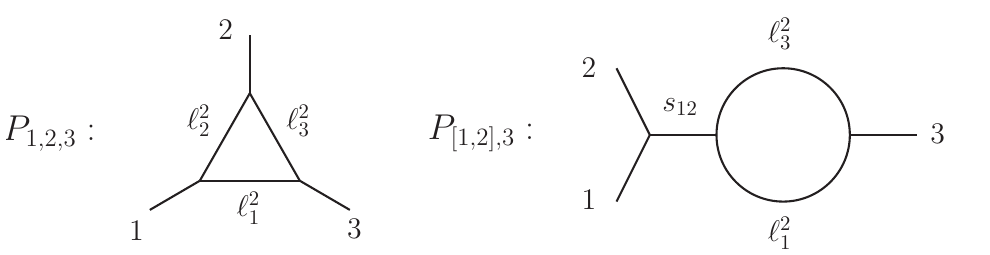}
    \captionsetup{justification=raggedright,singlelinecheck=false, margin=-1pt}
    \caption{Two diagrams contributing to $\mathcal{M}_{3}(1,2,3|1,2,3)$.}
    \label{fig_3pt_exp}
\end{figure}

For $n=3$, the kernel becomes $2\times2$,
\begin{equation}
\hspace{-0.6em}
  \begin{aligned}
 \mathcal{G}_3\!\!= \!\!
 \begin{pmatrix}
\mathcal{I}_{1,2,3}\!& \mathcal{I}_{1,3,2} 
\end{pmatrix}
\!\!
\left(\begin{smallmatrix}
\mathcal{M}_{3}(1,2,3|1,2,3) & \mathcal{M}_{3}(1,2,3|1,3,2)\\
 \mathcal{M}_{3}(1,3,2|1,2,3) & \mathcal{M}_{3}(1,3,2|1,3,2)
\end{smallmatrix}\right)^{-1}
\!
\begin{pmatrix}
\mathcal{I}_{1,2,3}\\
\mathcal{I}_{1,3,2}
\end{pmatrix}
\end{aligned}  
\end{equation}
where we have the shorthand notation $\mathcal{I}_{\alpha}:= \mathcal{I}_n(\alpha)$, and denote the poles of trivalent graphs $P_{1,2}=\ell_1^2 \ell_2^2$; the matrix elements are $\mathcal{M}_{3}(1,2,3|1,2,3)=\frac{1}{P_{1,2,3}}{+}\frac{1}{P_{[1,2],3}}{+}\frac{1}{P_{1,[2,3]}}{+}\frac{1}{P_{[3,1],2}}$, $\mathcal{M}_{3}(1,2,3|1,3,2)=\mathcal{M}_{3}(1,3,2|1,2,3)= {-}\frac{1}{P_{[1,2],3}}{-}\frac{1}{P_{1,[2,3]}}{-}\frac{1}{P_{[3,1],2}}$, and $\mathcal{M}_{3}(1,3,2|1,3,2)=\frac{1}{P_{1,3,2}}{+}\frac{1}{P_{[1,2],3}}{+}\frac{1}{P_{1,[2,3]}}{+}\frac{1}{P_{[3,1],2}}$ where $P_{1,2,3}=\ell_1^2 \ell_2^2 \ell_3^2$ denotes the triangle graph, while {\it e.g.} $P_{[1,2],3}=\ell_1^2 \ell_3^2 s_{1,2}$ denotes the bubble with a massive corner of legs $1,2$, depicted in the Fig.~\ref{fig_3pt_exp}.

\paragraph{BCJ double copy relations} 
In this setup, the numerators obtained from single-cut reconstruction can equally be generated via the one-loop KLT kernel \eqref{eq_Nexpand}, which confirms that they coincide with the expected one-loop BCJ numerators. Their cyclic symmetry can then be verified directly in this representation.
In this way, the gravity loop integrand takes the form 
\begin{equation}\label{eq:NNM}
 \mathcal{G}_n =\sum_{\alpha,\beta\in S_{n{-}1}} {{\rm N}}_{1,\alpha;\ell} {{\rm N}}_{1,\beta;\ell} \mathcal{M}_{n}(1,\alpha|1,\beta)\,.
\end{equation}
Recall that $\mathcal{M}_{n}(1,\alpha|1,\beta)$ denotes the intersection of one-loop cubic diagrams under two different orderings $(1,\alpha)$ and $(1,\beta)$~\cite{Feng:2022wee,Dong:2023stt}:
\begin{equation}
    \mathcal{M}_{n}(1,\alpha|1,\beta)=\operatorname{sgn}_\alpha^\beta \sum_{g \in T(1, \alpha) \cap T(1, \beta)} \frac{1}{P_g}\,,
\end{equation}
where $T(1, \alpha)$ denotes the collection of one-loop cubic diagrams (except for the tadpoles which we discard) in the ordering $(1, \alpha)$, and $P_g$'s are the propagators in the diagram $g$. The overall sign is fixed by the ordering~\cite{Feng:2022wee,Dong:2023stt}. For example, we expand~\eqref{eq:NNM} into diagrams to obtain the BCJ double copy relation for $\mathcal{G}_3$:
\begin{equation}
\begin{aligned}
  \mathcal{G}_3&=
  \frac{ {{\rm N}}_{1,2,3;\ell}^2}{P_{1,2,3}}{+}\frac{ {{\rm N}}_{1,3,2;\ell}^2}{P_{1,3,2}}{+}\frac{{{\rm N}}_{[1,2],3;\ell}^2}{P_{[1,2],3}}{+}\frac{ {{\rm N}}_{1,[2,3];\ell}^2}{P_{1,[2,3]}}{+}\frac{ {{\rm N}}_{[3,1],2;\ell}^2}{P_{[3,1],2}}\,.
\end{aligned}
\end{equation}

We emphasize that although the BCJ numerators obtained in this way may contain poles, they still yield correct YM/GR amplitudes.  There are redundancies which one can use to rewrite such numerators, and in practice it is straightforward to derive ``nice" BCJ numerators which are local ({\it i.e.} free of poles) and crossing symmetric. For $n=3,4,5$, using single cuts (from forward limits of up to seven-gluon trees), we obtain compact BCJ numerators, listed in the Appendix~\ref{app: bcjnum}.

\noindent {\bf Universal expansions for one-loop gravity integrands.}

We now apply the KLT relations to obtain a new representation of one-loop gravity integrands, as the analogue of the ``universal expansion'' for gauge-theory integrands. As shown in~\cite{Cao:2024olg}, a Yang-Mills integrand can be expanded in terms of gauge-invariant building blocks multiplied by mixed Yang-Mills-Scalar (YMS) loop integrands with scalars running in the loop,
\begin{equation}
{\cal I}_n^{\rm YM}(\mathbb{I})=\sum_{m;\alpha\in \tilde{S}_{m-1}} \Tr_\alpha {\cal I}^{\text{YMS}}_{\rm scalar-loop}(\alpha|\mathbb{I})\,,
\label{eq: I_n^YM expansion}
\end{equation}
where $m=0,\dots,n$, $\tilde{S}_{m-1}:= S_{m-1}/\mathbb{Z}_2$, and $\Tr_\alpha:=\Tr(f_{\alpha_{[1]}}\cdots f_{\alpha_{[m]}})$ with $f_i^{\mu\nu}=\epsilon_i^{[\mu} k_i^{\nu]}$. For $m=0$ we set $\Tr_\emptyset:= D{-}2$, where $D$ is the spacetime dimension. The object ${\cal I}^{\text{YMS}}_{\rm scalar-loop}(\alpha|\mathbb{I})$ denotes a YMS one-loop integrand with scalars in $\alpha$ and gluons in $\bar{\alpha}=\mathbb{I}/\alpha$ under ordering $\mathbb{I}$.

Substituting \eqref{eq: I_n^YM expansion} into \eqref{eq:one-loop KLT}, 
we obtain a universal expansion for the GR integrand:
\begin{equation}\label{eq: I_n^GR expansion}
\begin{aligned}
    {\cal G}_n^{\rm GR}={\sum_{\substack{m;\alpha\in \tilde{S}_{m{-}1}} }}{\sum_{\substack{p;\beta\in \tilde{S}_{p{-}1}}} }\Tr_\alpha \Tr_\beta {\cal G}^{\text{EYMS}}_{\rm scalar-loop}(s|g|h)\,,    
\end{aligned}
\end{equation}
where ${\cal G}^{\text{EYMS}}_{\rm scalar-loop}$ denotes the one-loop integrands of Einstein-Yang-Mills-Scalars theory with scalar running in the loop~\cite{Porkert:2022efy}, and we have scalars in $\alpha \cap\beta$ $(s)$, gluons in $(\alpha \cap\bar{\beta})\cup(\bar{\alpha} \cap\beta)$ $(g)$, and gravitons in $\bar{\alpha} \cap\bar\beta$ $(h)$. These integrands can themselves be constructed from YMS scalar-loop integrands through a one-loop KLT formula:
\begin{equation}\label{eq:EYMS KLT}
\begin{aligned}
    &{\cal G}^{\text{EYMS}}_{\rm scalar-loop}(\alpha \cap\beta|(\alpha \cap\bar{\beta})\cup(\bar{\alpha} \cap\beta)|\bar{\alpha} \cap\bar\beta) \\
    &=\!\!\!\!\sum_{\gamma,\delta\in S_{n{-}1}} \!\!\!\!\mathcal{I}^\text{YMS}_{\rm scalar-loop}(\alpha|1,\gamma) \mathcal{K}(1,\gamma|1,\delta)\mathcal{I}^\text{YMS}_{\rm scalar-loop}(\beta|1,\delta) 
\end{aligned}.
\end{equation}

As a simple example, consider the one-loop two-graviton  integrand, which expands as 
${\cal G}_2^{\rm GR}{=}\Tr_\emptyset^2{\cal G}(1^h,2^h){+}2\Tr_\emptyset\Tr_{1,2}{\cal G}(1^g,2^g){+}\Tr_{1,2}^2{\cal G}(1^s,2^s)$. This leads to 
\begin{align}
    {\cal G}_2^{\rm GR}{=} & {\Tr_{\emptyset}^2{\cal I}(1^g,2^g)^2}{P_{1,2}}{+}2{\Tr_{\emptyset}\Tr_{1,2}{\cal I}(1^g,2^g) {\cal I}(1^s,2^s)}{P_{1,2}}
    \nonumber
    \\ & {+}{\Tr_{1,2}^2{\cal I}(1^s,2^s)^2}{P_{1,2}}\,,
\end{align}
via KLT relations \eqref{eq:EYMS KLT}. It finally reduces to a compact form  ${\cal G}_2^{\rm GR}{=}{(\Tr_{\emptyset}\epsilon_{1}\cdot \ell_1 \epsilon_{2}\cdot \ell_2{+}\Tr_{1,2})^{2}}/{P_{1,2}}$,
 which agrees with the one-loop double-copy result with BCJ numerator ${\rm N}_{1,2;\ell}^{\text{YM}}=\Tr_\emptyset \epsilon_{1}\cdot \ell_1 \epsilon_{2}\cdot \ell_2{+}\Tr_{1,2}$.  Here we omit the superscript text in  $\mathcal{G/I}$  for shorthands.

Finally, we note that the universal expansion of~\cite{Cao:2024olg} extends to general gauge theories, including supersymmetric cases. In such theories, one encounters gluon, fermion, and scalar loops. The scalar-loop YMS integrands remain unchanged, while the trace structures $\Tr$ must be generalized to include not only vector but also spinor traces. The gravity analogue \eqref{eq: I_n^GR expansion} admits the same extension, yielding analogous expansions for one-loop graviton integrands in supergravity theories.

\noindent {\bf Outlook.}

In this letter, we have shown how to establish one-loop KLT and BCJ double copy relations by reconstructing one-loop integrands from forward limits of tree amplitudes. As long as there exist tree-level kinematic numerators whose forward-limits satisfy \eqref{cyclic-cons}, our method immediately leads to (a). a new way of extracting one-loop BCJ numerators from (forward limits of) trees, (b). one-loop KLT relations which do not need explicit BCJ numerators, \eqref{eq:one-loop KLT}, and as an application also (c). universal expansions of (super-)gravity amplitudes, which are analogous to that for (super-)Yang-Mills amplitude, \eqref{eq: I_n^GR expansion}. As we show in the Appendix~\ref{app: chy}, it is straightforward to derive one-loop double copy from one-loop CHY formulas, which provides an alternative perspective; we also provide compact results for Yang-Mills BCJ numerators up to five points, which are local and crossing symmetric.

Our preliminary investigations open up new avenues for future studies. To have a rigorous proof of one-loop double copy relations, the only missing ingredient seems to be a proof of the shifted cyclic symmetry, which can be viewed as certain compatibility conditions between forward-limits of tree-level kinematic numerators and the ``locality'' (no spurious poles) for one-loop integrands. This may shed new light on the origin of the so-called ``kinematic algebras"~\cite{Monteiro:2011pc, Cheung:2021zvb} beyond tree level, and relatedly it would be highly desirable to develop a direct method for computing all-multiplicity BCJ numerators at one loop level for (super-)YM/GR, extending results here and those in~\cite{Edison:2022jln, Dong:2023stt}. 

Our one-loop double copy relations should be closely related to the KLT relations for certain loop-integrand version of the open- and closed-string amplitudes at genus one~\cite{Mafra:2017ioj,Stieberger:2022lss,Stieberger:2023nol}, generalizing the beautiful connection at tree level. It would be extremely interesting to understand this completely, and make connections with results in~\cite{Balli:2024wje,Geyer:2024oeu, Geyer:2021oox}. On the other hand, regardless of double copy, our way for constructing the gravity loop integrands should have a wider range of implications, and it also sheds new light on how to produce gravity loop amplitudes from ``surfaceology". 

Most importantly, can we extend our method to higher loops? In~\cite{Cao:2025mlt}, single-cut reconstruction has been realized for higher-loop (planar) integrands in gauge theories, and at least two ingredients are needed to apply this to gravity as the double copy at higher loops: a basis of bi-adjoint amplitudes built from forward limits of lower-loop ones (which we conjecture to have rank $(n{-}3{+}2L)!$ at $L$ loops), as well as generalizations of the shifted cyclic symmetry to higher loops. We leave the study of these exciting questions to future works.

\noindent {\bf Acknowledgements.---}
We thank Jin Dong, Yi-Jian Du, and Chongsi Xie for discussions. The work of Q.C. is supported by the National Natural Science Foundation of China under Grant No. 123B2075 and No. 12347103.  The work of S.H. has been supported by the National Natural Science Foundation of China under Grant No. 12225510, 12047503, 12247103, and by the New Cornerstone Science Foundation through the XPLORER PRIZE.
The work of Y.Z. is supported by the National Natural Science Foundation of China under Grant No. 12405086.
The work of F.Z. is supported in part by the Science Challenge Project (No. TZ2025012), and NSAF No. U2330401.

\onecolumngrid

\appendix
\section{Examples of one-loop YM/GR integrands from forward limits
}\label{app:example}

Here we illustrate the reconstruction of one-loop integrands in Yang--Mills theory and in gravity based on single cuts, obtained from forward limits of trees. 

\subsection{Yang-Mills loop integrands}
We start with the simplest example: the two-point one-loop Yang--Mills integrand. According to~\eqref{eq:I-singlecut}, it takes the form
\begin{equation}\label{eq:2pt}
   {\tilde {\mathcal{I}}}_{2}(1,2)
   =  \frac{\hat{\mathcal{A}}_{4}(+,1,2,-)}{\ell_{1}^{2}}\,+ \frac{\hat{\mathcal{A}}_{4}(+,2,1,-)}{\ell_{2}^{2}}\,,
\end{equation}
where $\hat{\mathcal A}$ denotes the forward limit of a tree amplitude.  

Expanding the two terms in the $m_4$ basis, one finds
\begin{equation}
\begin{aligned}
\begin{pmatrix}
 \frac{\hat{\mathcal{A}}_{4}(+,1,2,-)}{\ell_{1}^{2}}\\
\frac{\hat{\mathcal{A}}_{4}(+,2,1,-)}{\ell_{2}^{2}}
\end{pmatrix}=& \begin{pmatrix}
 \frac{{\rm N}_{1,2;\ell_1} \,m_{4}(+,1,2,-\,|\, +,1,2,-)}{\ell_1^2} 
    +  \frac{{\rm N}_{2,1;\ell_1} \,
     m_{4}(+,1,2,-\,|\, +,2,1,-)}{\ell_1^2} \\
  \frac{{\rm N}_{2,1;\ell_2} \,
     m_{4}(+,2,1,-\,|\, +,2,1,-)}{\ell_2^2}    +  \frac{{\rm N}_{1,2;\ell_2} \,
     m_{4}(+,2,1,-\,|\, +,1,2,-)}{\ell_2^2} 
\end{pmatrix}\\
=& \begin{pmatrix}
 {\rm N}_{1,2;\ell_1} (\frac{1}{Y_1 (Y_2-Y_1)}+\frac{1}{Y_1 X_{1,1}})
    +  {\rm N}_{2,1;\ell_1}(-\frac{1}{Y_1 X_{1,1}})\\
 {\rm N}_{2,1;\ell_2}(\frac{1}{Y_2 (Y_1-Y_2)}+\frac{1}{Y_2 X_{2,2}})   + {\rm N}_{1,2;\ell_2}(-\frac{1}{Y_2 X_{2,2}})
\end{pmatrix}\,,
\end{aligned}
\end{equation}
where $\ell_1^2=Y_1$ and $\ell_2^2=Y_2$.  

Cancellation of the spurious pole $(Y_1-Y_2)$ requires
$
{\rm N}_{1,2;\ell_1}={\rm N}_{2,1;\ell_2},
$
which is exactly the shifted cyclic symmetry~\eqref{cyclic-cons}.  
Together with the crossing relation ${\rm N}_{2,1;\ell_1}={\rm N}_{1,2;\ell_1}|_{1\leftrightarrow 2}$, this further implies ${\rm N}_{2,1;\ell_1}={\rm N}_{1,2;\ell_2}$.  

Summing the two contributions, the integrand can be written in a $2!$ basis:
\begin{equation}
\begin{aligned}
  {\tilde {\mathcal{I}}}_{2}(1,2)
   &= {\rm N}_{1,2;\ell_1} (\frac{1}{Y_1 Y_2}+\frac{1}{Y_1 X_{1,1}}+\frac{1}{Y_2 X_{2,2}})+ {\rm N}_{2,1;\ell_1} (-\frac{1}{Y_1 X_{1,1}}-\frac{1}{Y_2 X_{2,2}})\\
   &=:  {\rm N}_{1,2;\ell_1} \mathcal{M'}(1,2|1,2)+ {\rm N}_{2,1;\ell_1} \mathcal{M'}(1,2|2,1)\,.
\end{aligned}
\end{equation}
Although $\mathcal{M}'$ involves only quadratic propagators, it is over-complete. 
By cyclic symmetry one has 
$
{\rm N}_{2,1;\ell_1}={\rm N}_{1,2;\ell_2}$ with $ \ell_2=\ell+k_1,
$
and after shifting $\ell_2\to \ell_1$ this reduces to 
$
{\rm N}_{2,1;\ell_1}\big|_{\ell_2\to\ell_1}={\rm N}_{1,2;\ell_1}\,.
$
Thus the expression can be reduced to the minimal $1!$ basis:
\begin{equation}
\begin{aligned}
  \mathcal{I}_{2}(1,2)
   &= {\rm N}_{1,2;\ell_1} (\frac{1}{Y_1 Y_2})=:  {\rm N}_{1,2;\ell_1} \mathcal{M}(1,2|1,2)\cong {\rm N}_{1,2;\ell_1}(\mathcal{M'}(1,2|1,2)+\mathcal{M'}(1,2|2,1)\big|_{\ell_2\to\ell_1})
\end{aligned}\,.
\end{equation}

Let us now turn to the three-point example,
\begin{equation}\label{eq:3pt}
   {\tilde {\mathcal{I}}}_{3}(1,2,3)
   =  \frac{\hat{\mathcal{A}}_{5}(+,1,2,3,-)}{\ell_{1}^{2}}\,+ \frac{\hat{\mathcal{A}}_{5}(+,2,3,1,-)}{\ell_{2}^{2}}+\frac{\hat{\mathcal{A}}_{5}(+,3,1,2,-)}{\ell_{3}^{2}}\,,
\end{equation}
and expand each term in the $m_5$ basis with propagators linear in $\ell$:
\begin{align}
&
\begin{pmatrix}
 \frac{\hat{\mathcal{A}}_{5}(+,1,2,3,-)}{\ell_{1}^{2}}\\
\frac{\hat{\mathcal{A}}_{5}(+,2,3,1,-)}{\ell_{2}^{2}}\\
\frac{\hat{\mathcal{A}}_{5}(+,3,1,2,-)}{\ell_{3}^{2}}
\end{pmatrix}= 
\nonumber
\\ &
\begin{aligned}
& \begin{pmatrix}
 \frac{{\rm N}_{1,2,3;\ell_1} \,m_{5}(a|1,2,3)}{\ell_1^2} 
  \!+\!  \frac{{\rm N}_{1,3,2;\ell_1} \,
     m_{5}(a|1,3,2)}{\ell_1^2}  \!+\!\frac{{\rm N}_{2,1,3;\ell_1} \,
     m_{5}(a|2,1,3)}{\ell_1^2} \!+\!\frac{{\rm N}_{2,3,1;\ell_1} \,
     m_{5}(a|2,3,1)}{\ell_1^2} \!+\!\frac{{\rm N}_{3,1,2;\ell_1} \,
     m_{5}(a|3,1,2)}{\ell_1^2} \!+\!\frac{{\rm N}_{3,2,1;\ell_1} \,
     m_{5}(a|3,2,1)}{\ell_1^2}\\
 \frac{{\rm N}_{2,3,1;\ell_2} \,m_{5}(b|2,3,1)}{\ell_2^2} 
   \!+\!  \frac{{\rm N}_{3,2,1;\ell_2} \,
     m_{5}(b|3,2,1)}{\ell_2^2}  \!+\!\frac{{\rm N}_{1,3,2;\ell_2} \,
     m_{5}(b|1,3,2)}{\ell_2^2} \!+\!\frac{{\rm N}_{3,1,2;\ell_2} \,
     m_{5}(b|3,1,2)}{\ell_2^2} \!+\!\frac{{\rm N}_{1,2,3;\ell_2} \,
     m_{5}(b|1,2,3)}{\ell_2^2} \!+\!\frac{{\rm N}_{2,1,3;\ell_2} \,
     m_{5}(b|2,1,3)}{\ell_2^2}\\
     \frac{{\rm N}_{3,1,2;\ell_3} \,m_{5}(c|3,1,2)}{\ell_3^2} 
   \!+\!  \frac{{\rm N}_{2,1,3;\ell_3} \,
     m_{5}(c|2,1,3)}{\ell_3^2}  \!+\!\frac{{\rm N}_{3,2,1;\ell_3} \,
     m_{5}(c|3,2,1)}{\ell_3^2} \!+\!\frac{{\rm N}_{1,2,3;\ell_3} \,
     m_{5}(c|1,2,3)}{\ell_3^2} \!+\!\frac{{\rm N}_{2,3,1;\ell_3} \,
     m_{5}(c|2,3,1)}{\ell_1^2} \!+\!\frac{{\rm N}_{1,3,2;\ell_3} \,
     m_{5}(c|1,3,2)}{\ell_3^2}
\end{pmatrix},
\end{aligned}
\end{align}
where $a=(+,1,2,3,-)$, $b=(+,2,3,1,-)$, $c=(+,3,1,2,-)$, and for brevity we suppress the external $+,-$ labels in $m_5$.

It is convenient to reorganize the terms row by row, so that columns are grouped according to the shifted cyclic symmetry of the numerators~\eqref{cyclic-cons}. The consistency conditions then follow from demanding that all spurious poles cancel in the final result. For instance, focusing on the first column and substituting the explicit form of $m_5$, one encounters half-ladder diagrams with linear propagators,
\begin{equation}\label{eq:N3cyc}
\begin{pmatrix}
\frac{{\rm N}_{1,2,3;\ell_1} \,m_{5}(a|1,2,3)}{\ell_1^2}\\
\frac{{\rm N}_{2,3,1;\ell_2} \,m_{5}(b|2,3,1)}{\ell_2^2}\\
\frac{{\rm N}_{3,1,2;\ell_3} \,m_{5}(c|3,1,2)}{\ell_3^2}
\end{pmatrix}=   \begin{pmatrix}
{\rm N}_{1,2,3;\ell_1} (\frac{ 1}{Y_1 \left(Y_1-Y_2\right) \left(Y_1-Y_3\right)} +\cdots)\\
{\rm N}_{2,3,1;\ell_2} (\frac{ 1}{Y_2 \left(Y_2-Y_1\right) \left(Y_2-Y_3\right)} +\cdots)\\
{\rm N}_{3,1,2;\ell_3} (\frac{ 1}{Y_3 \left(Y_3-Y_1\right) \left(Y_3-Y_2\right)} +\cdots)\end{pmatrix}.
\end{equation}
 Their denominators contain factors such as $(Y_1{-}Y_2)$, $(Y_1{-}Y_3)$, and $(Y_2{-}Y_3)$, and the absence of these spurious poles requires the corresponding numerators to be identified across the three rows,  ${\rm N}_{1,2,3;\ell_1}={\rm N}_{2,3,1;\ell_2}={\rm N}_{3,1,2;\ell_3}$.  The three contributions then combine into the triangle diagram,  ${\rm N}_{1,2,3;\ell_1} (\frac{ 1}{Y_1 \left(Y_1-Y_2\right) \left(Y_1-Y_3\right)}+\frac{ 1}{Y_2 \left(Y_2-Y_1\right) \left(Y_2-Y_3\right)}+\frac{ 1}{Y_3 \left(Y_3-Y_1\right) \left(Y_3-Y_2\right)})= \frac{{\rm N}_{1,2,3;\ell_1}}{Y_1 Y_2Y_3}$. 
 Including the crossing relations among tree-level BCJ numerators, all numerators in each column can be identified. This leads to the intermediate $3!=6$ basis $\mathcal{M}'_3(\mathbb{I}|\cdots)$, listed as follows,
\begin{equation}   
\begin{pmatrix}
     \{1,2,3\} & \frac{1}{Y_1 Y_2 X_{1,2}}{+}\frac{1}{Y_1 X_{1,1} X_{1,2}}{+}\frac{1}{Y_1 Y_3 X_{1,3}}{+}\frac{1}{Y_1 X_{1,1} X_{1,3}}{+}\frac{1}{Y_2 X_{1,2} X_{2,2}}{+}\frac{1}{Y_2 Y_3 X_{2,3}}{+}\frac{1}{Y_2 X_{2,2} X_{2,3}}{+}\frac{1}{Y_3 X_{1,3} X_{3,3}}{+}\frac{1}{Y_3 X_{2,3} X_{3,3}}{+}\frac{1}{Y_1 Y_2 Y_3} \\
 \{1,3,2\} & -\frac{1}{Y_1 X_{1,1} X_{1,2}}-\frac{1}{Y_2 X_{1,2} X_{2,2}}+\frac{1}{Y_3 X_{1,3} X_{3,3}}+\frac{1}{Y_3 X_{2,3} X_{3,3}}-\frac{1}{Y_1 Y_2 X_{1,2}} \\
 \{2,1,3\} & -\frac{1}{Y_1 X_{1,1} X_{1,3}}+\frac{1}{Y_2 X_{1,2} X_{2,2}}+\frac{1}{Y_2 X_{2,2} X_{2,3}}-\frac{1}{Y_3 X_{1,3} X_{3,3}}-\frac{1}{Y_1 Y_3 X_{1,3}} \\
 \{2,3,1\} & -\frac{1}{Y_2 X_{2,2} X_{2,3}}-\frac{1}{Y_3 X_{1,3} X_{3,3}}-\frac{1}{Y_1 X_{1,1} X_{1,2}} \\
 \{3,1,2\} & -\frac{1}{Y_2 X_{1,2} X_{2,2}}-\frac{1}{Y_3 X_{2,3} X_{3,3}}-\frac{1}{Y_1 X_{1,1} X_{1,3}} \\
 \{3,2,1\} & \frac{1}{Y_1 X_{1,1} X_{1,3}}-\frac{1}{Y_2 Y_3 X_{2,3}}-\frac{1}{Y_2 X_{2,2} X_{2,3}}-\frac{1}{Y_3 X_{2,3} X_{3,3}}+\frac{1}{Y_1 X_{1,1} X_{1,2}} 
\end{pmatrix}.
\end{equation}

We emphasize that $\mathcal{M}'$ is over-complete and only serves as an intermediate basis. Using cyclic symmetry and loop-momentum shifts, e.g.
${\rm N}_{2,3,1;\ell_1}={\rm N}_{1,2,3;\ell-k_1}\to{\rm N}_{1,2,3;\ell}$ and ${\rm N}_{3,1,2;\ell_1}={\rm N}_{1,2,3;\ell-(k_1+k_2)}\to{\rm N}_{1,2,3;\ell}$, one can reduce to the minimal $2!$ basis $\mathcal{M}_3$ given in~\eqref{eq:M},
\begin{equation}
  \mathcal{M}(1,2,3|1,2,3)=\frac{1}{Y_1Y_2Y_3}+\frac{1}{Y_1Y_2X_{1,2}}+\frac{1}{Y_1Y_3X_{1,3}}+\frac{1}{Y_2Y_3X_{2,3}},\quad   \mathcal{M}(1,2,3|1,3,2)=-\frac{1}{Y_1Y_2X_{1,2}}-\frac{1}{Y_1Y_3X_{1,3}}-\frac{1}{Y_2Y_3X_{2,3}}\,.
\end{equation}
Finally, the three-point one-loop Yang-Mills integrand is
\begin{equation}
  \mathcal{I}_{3}(1,2,3)
  = {\rm N}_{1,2,3;\ell_1} \mathcal{M}(1,2,3|1,2,3)+ {\rm N}_{1,3,2;\ell_1} \mathcal{M}(1,2,3|1,3,2)\,.
\end{equation}

\subsection{Gravity loop integrands}

To construct gravity integrands we make use of the partial integrands $G_n(\alpha)$ introduced in the main text, cf.~\eqref{condition1}--\eqref{condition2}.
These objects serve as ordered building blocks: although gravity amplitudes themselves are orderless, assigning an auxiliary ordering label $\alpha$ allows us to apply the same single-cut reconstruction used for Yang--Mills. Summing over $\alpha$ restores the full gravity integrand $\tilde{\mathcal G}_n$.

For concreteness, let us illustrate how this works for $n=2$ and $n=3$.

The simplest case is the two-point one-loop gravity integrand. According to the definition of the partial integrands $G_n(\alpha)$, the full integrand is obtained by summing over both orderings,
\begin{equation}
 \tilde {\mathcal{G}}_2=G_2(1,2)+G_{2}(2,1)\,.
\end{equation}

Let us first consider the canonical ordering $(1,2)$.
Using the residue-theorem representation~\eqref{residueG}, we shift $Y_i\to Y_i -z$ and evaluate at $z=0$,
\begin{equation}
\begin{aligned}
  G_2(1,2)&=  
    \frac{1}{Y_1} \mathrm{Res}_{Y_1=0}G_{2}(1,2)\Big|_{Y_2\to Y_2-Y_1}+\frac{1}{Y_2} \mathrm{Res}_{Y_2=0}G_{2}(1,2)\Big|_{Y_1\to Y_1-Y_2}\\
    &= {\rm N}_{1,2;\ell_1}\frac{\hat{\mathcal{A}}_{4}(+,1,2,-)}{\ell_{1}^{2}} +{\rm N}_{2,1;\ell_2}\frac{\hat{\mathcal{A}}_{4}(+,2,1,-)}{\ell_{2}^{2}}\,,
\end{aligned}    
\end{equation}
where condition~\eqref{condition2} has been used to substitute the residues.

As in the Yang--Mills case, cancellation of spurious poles requires the shifted cyclic symmetry ${\rm N}_{1,2;\ell_1}={\rm N}_{2,1;\ell_2}$. The two contributions can then be combined,
\begin{equation}
\begin{aligned}
  G_2(1,2)
    &= {\rm N}_{1,2;\ell_1}(\frac{\hat{\mathcal{A}}_{4}(+,1,2,-)}{\ell_{1}^{2}} +\frac{\hat{\mathcal{A}}_{4}(+,2,1,-)}{\ell_{2}^{2}})\\
    &={\rm N}_{1,2;\ell_1} \tilde{\mathcal{I}}_2(1,2)\,,
\end{aligned}    
\end{equation}
where we used the two-point Yang--Mills integrand~\eqref{eq:2pt}.

The other ordering can be obtained either by repeating the same steps or simply by relabeling, $G_2(2,1)=G_2(1,2)\big|_{1\leftrightarrow2}={\rm N}_{2,1;\ell_1} \tilde{\mathcal{I}}_2(2,1)$ with the loop momentum left unchanged. 

Finally, shifted cyclic symmetry implies $G_2(1,2)\cong G_2(2,1)$, and the full integrand becomes  
\begin{equation}
     \tilde {\mathcal{G}}_2=G_2(1,2)+G_2(2,1)
     \cong 2G_2(1,2)\,.
\end{equation}

Let us now turn to the three-point one-loop case,
\begin{equation}
 \tilde {\mathcal{G}}_3=G_3(1,2,3)+G_{3}(1,3,2)+G_3(2,1,3)+G_{3}(2,3,1)+G_3(3,1,2)+G_{3}(3,2,1)\,.
\end{equation}

Consider first the canonical ordering $(1,2,3)$.
Applying the residue-theorem construction~\eqref{residueG}, one finds
\begin{equation}
\begin{aligned}
  G_3(1,2,3)&=  
    \frac{1}{Y_1} \mathrm{Res}_{Y_1=0}G_{3}(1,2,3)\Big|_{Y_j\to Y_j-Y_1}+\frac{1}{Y_2} \mathrm{Res}_{Y_2=0}G_{3}(1,2,3)\Big|_{Y_j\to Y_j-Y_2}+\frac{1}{Y_3} \mathrm{Res}_{Y_3=0}G_{3}(1,2,3)\Big|_{Y_j\to Y_j-Y_3}\\
    &= {\rm N}_{1,2,3;\ell_1}\frac{\hat{\mathcal{A}}_{5}(+,1,2,3,-)}{\ell_{1}^{2}} +{\rm N}_{2,3,1;\ell_2}\frac{\hat{\mathcal{A}}_{5}(+,2,3,1,-)}{\ell_{2}^{2}}+{\rm N}_{3,1,2;\ell_3}\frac{\hat{\mathcal{A}}_{5}(+,3,1,2,-)}{\ell_{3}^{2}}\\
    &={\rm N}_{1,2,3;\ell_1}(\frac{\hat{\mathcal{A}}_{5}(+,1,2,3,-)}{\ell_{1}^{2}} +\frac{\hat{\mathcal{A}}_{5}(+,2,3,1,-)}{\ell_{2}^{2}}+\frac{\hat{\mathcal{A}}_{5}(+,3,1,2,-)}{\ell_{3}^{2}})\\
    &={\rm N}_{1,2,3;\ell_1}\tilde{\mathcal{I}}_3(1,2,3)\,,
\end{aligned}    
\end{equation}
where shifted cyclic symmetry for the numerators~\eqref{cyclic-cons} (see also~\eqref{eq:N3cyc}) has been used, together with the three-point Yang--Mills integrand~\eqref{eq:3pt}. 

Other ordered partial integrands $G_3(\alpha)$ can either be obtained in the same way or simply by relabeling. Taking cyclic symmetry into account, the final result simplifies to
\begin{equation}
 \tilde {\mathcal{G}}_3\cong 3G_3(1,2,3)+3G_{3}(1,3,2)=3({\rm N}_{1,2,3;\ell_1}\tilde{\mathcal{I}}_3(1,2,3)+{\rm N}_{1,3,2;\ell_1}\tilde{\mathcal{I}}_3(1,3,2))\,,
\end{equation}
where we used, for example, that $G_3(2,3,1)={\rm N}_{2,3,1;\ell_1}\tilde{\mathcal{I}}_3(2,3,1)={\rm N}_{1,2,3;\ell_1-k_1}\tilde{\mathcal{I}}_3(1,2,3)={\rm N}_{1,2,3;\ell_1}\tilde{\mathcal{I}}_3(1,2,3)\big|_{\ell_1\to\ell_1-k_1}\cong {\rm N}_{1,2,3;\ell_1}\tilde{\mathcal{I}}_3(1,2,3)=G_3(1,2,3)$.

\section{Local, crossing symmetric BCJ numerators for YM/GR}\label{app: bcjnum}
In this appendix we present explicit local BCJ numerators for pure Yang--Mills amplitudes at one loop, up to five external legs. These numerators are polynomial in Lorentz products of the loop and external momenta as well as the polarization vectors, and are given in Eqs.~\eqref{eq_3ptbcj}--\eqref{eq_5ptbcj}.  

\begin{figure}[htbp]
    \centering
    \includegraphics[width=0.7\linewidth]{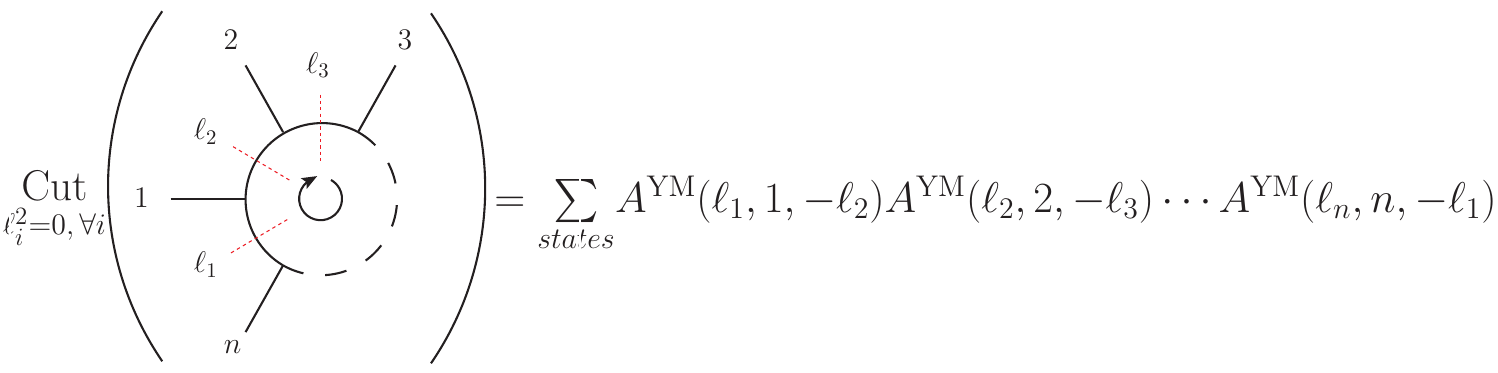}
    \captionsetup{justification=raggedright,singlelinecheck=false}
    \caption{Maximal-cut numerator for the $n$-gon. The arrow on the left indicates the orientation of the loop momentum. One can easily prove that the  right-hand side is exactly~\eqref{eq_maximalnum}.}
    \label{fig_maximalnum}
\end{figure}

To set up the results, it is convenient to introduce the \emph{maximal-cut numerator}, shown in Fig.~\ref{fig_maximalnum}, which naturally appears as a component of the full BCJ numerator. For an $n$-point one-loop amplitude it is defined as
\begin{equation}
C_{12\ldots n}^{\mathrm{YM}}:=\Tr(g_1g_2\ldots g_n),\qquad {\rm with} ~g_i^{\mu\nu}:=f_i^{\mu\nu}-(\ell_i\cdot \epsilon_i)\eta^{\mu\nu}\,.
\label{eq_maximalnum}
\end{equation}
As a simple example, $C_{1}^{\rm YM}=-(D-2)\ell_1\cdot\epsilon_1 $, $C_{12}^{\rm YM}=\Tr(f_1f_2)+(D-2)\ell_1\cdot\epsilon_1 \ell_2\cdot\epsilon_2$, where $D$ denotes the spacetime dimension. For completeness we also define $C_{\varnothing}^{\rm YM}:=\Tr(\emptyset)=D-2$.

As it turns out, the space of one-loop $n$-point BCJ numerators with crossing symmetry is highly non-unique. For simplicity, here we present one particular solution that yields relatively compact expressions.

For $n=3,4$, the numerators can be neatly expressed in terms of maximal-cut numerators, supplemented by simple factors such as $\epsilon_i\cdot \epsilon_j$ and $Y_i=\ell_i^2$:
\begin{equation}
	{\rm N}_{123}^{\rm{YM}}=C^{\mathrm{YM}}_{123}-\frac{1}{4}\left(C^{\mathrm{YM}}_{1}\epsilon_2\cdot\epsilon_3 Y_3+\textit{cyclic.}\right)
    \label{eq_3ptbcj}
\end{equation}

\begin{equation}
	{\rm N}_{1234}^{\text{YM}}=C^{\mathrm{YM}}_{1234}-\frac{1}{4}\left(C^{\mathrm{YM}}_{12}\epsilon_3\cdot\epsilon_4 Y_4-\frac{1}{8}C_{\varnothing}^{\rm YM}\epsilon_1\cdot\epsilon_2\epsilon_3\cdot\epsilon_4 Y_2 Y_4+\textit{cyclic.}\right)
    \label{eq_4ptbcj}
\end{equation}
At five points the structure becomes more involved. A representative solution can be written as
\begin{equation}
	{\rm N}_{12345}^{\text{YM}}=C^{\mathrm{YM}}_{12345}-\frac{1}{4}\left(C'_{123}\epsilon_4\cdot\epsilon_5 Y_5-\frac{1}{4}C'_{1}\epsilon_2\cdot\epsilon_3 \epsilon_4\cdot\epsilon_5Y_3 Y_5+\textit{cyclic.}\right)
    \label{eq_5ptbcj}
\end{equation}
where we have introduced the ``deformed'' cuts $C'_1$ and $C'_{123}$, defined as
\begin{equation}
	C'_{1}=C_{1}^{\rm YM}-C_{\varnothing}^{\rm YM}(k_3+k_5)\cdot\epsilon_1,
\end{equation}
and
\begin{equation}
	\begin{aligned}
		C'_{123}=&\cym_{123}-\Tr(f_1f_2)(\frac{k_5-k_4}{2})\cdot\epsilon_3-\Tr(f_2f_3)(\frac{k_5-k_4}{2})\cdot\epsilon_1+\Tr(f_1f_3)(\frac{k_5-k_4}{2})\cdot\epsilon_2\\
		&+\frac{C_{\varnothing}^{\rm YM}}{4}\Bigg[2 \ell_2\cdot \epsilon _2 \left(\ell_4+\ell_5-\ell_1\right)\cdot \epsilon _1 \ell_3\cdot \epsilon _3+\ell_1\cdot \epsilon _1 \left(-\ell_1-\ell_4+2 \ell_5\right)\cdot \epsilon _2 \ell_3\cdot \epsilon _3\\
		&+\left(\ell_1+\ell_4\right)\cdot \epsilon _2 \left(\ell_4-2 \ell_5\right)\cdot \epsilon _1 \ell_3\cdot \epsilon _3-2 \ell_2\cdot \epsilon _2 \ell_4\cdot \epsilon _1 \ell_5\cdot \epsilon _3\\
		&+2 \ell_1\cdot \epsilon _1 \ell_2\cdot \epsilon _2 \left(\ell_1+\ell_5\right)\cdot \epsilon _3+2 \ell_1\cdot \epsilon _3 \ell_2\cdot \epsilon _2 \left(\ell_4+\ell_5\right)\cdot \epsilon _1-\ell_1\cdot \epsilon _1 \left(\ell_1+\ell_4\right)\cdot \epsilon _2 \left(\ell_1-2 \ell_5\right)\cdot \epsilon _3\\
		&-\ell_1\cdot \epsilon _3 \ell_4\cdot \epsilon _1 \left(\ell_1+\ell_4-2 \ell_5\right)\cdot \epsilon _2\Bigg]\,.
	\end{aligned}
\end{equation}

\section{Relations with one-loop CHY formulas}\label{app: chy}
In this appendix we review the one-loop CHY formulation and its connection to our results. We show that the one-loop KLT relations here differ from the linear-propagator version by a basis transformation, and explain how generalized Parke--Taylor factors yield \eqref{grint}, \eqref{eq:one-loop KLT} and \eqref{eq:NNM} directly from \eqref{eq_Iexpand}. 

The one-loop CHY integrand can be written as the inner product of two half integrands defined on the $n$-punctured Riemann sphere with loop momentum $\ell$ \cite{He:2015yua,Cachazo:2015aol},
\begin{equation}
\langle \mathcal{I}_L|\mathcal{I}_R\rangle := \frac{1}{\ell^2} \int \underbrace{\prod_{i=2}^n \mathrm{~d} \sigma_i \,\delta\!\left(\frac{2 \ell \cdot k_i}{\sigma_i}+\sum_{\substack{j=1 \\ j \neq i}}^n \frac{s_{i j}}{\sigma_{i j}}\right)}_{:= \mathrm{d} \mu_n} \,\mathcal{I}_L(\ell)\,\mathcal{I}_R(\ell)\,,
\end{equation}
where $ \sigma_{ij}:=\sigma_i-\sigma_j$ and $s_{ij}:=2k_i\cdot k_j$. For convenience we fix $\sigma_1=0$. These formulas are supported on the one-loop scattering equations, which arise from the forward limit of $(n{+}2)$-point tree-level equations. We focus on Yang--Mills and gravity, for which the relevant half integrands are
\begin{equation}
\begin{aligned}
& \mathrm{PT}(1,2,\ldots,n):= \sum_{i=1}^n \mathrm{PT}^{\rm tree}(+,i,i{+}1,\ldots,i{-}1,-)\,,\\
& \mathrm{Pf}(\{1,2,\ldots,n\}):= \hat{\mathrm{Pf}}^{\rm tree}(\{+,1,2,\ldots,n,-\})\,,
\end{aligned}
\end{equation}
where the one-loop Parke--Taylor factor is a cyclic sum over tree-level factors, and the one-loop Pfaffian is obtained from the forward limit of the $(n{+}2)$-point tree-level Pfaffian.

Their inner products generate the one-loop bi-adjoint, Yang--Mills and gravity integrands,
\begin{equation}
\begin{aligned}
     &\langle \mathrm{PT}(\alpha)|\mathrm{PT}(\beta)\rangle=\mathcal{M}(\alpha|\beta)\,,\qquad 
     \langle \mathrm{Pf}|\mathrm{PT}(\alpha)\rangle=\mathcal{I}^{\text{YM}}(\alpha)\,,\qquad 
     \langle \mathrm{Pf}|\mathrm{Pf}\rangle=\mathcal{G}^{\text{GR}}_n\,.
\end{aligned}
\end{equation}

\subsection{KLT double copy}

By construction, gravity integrands arise as the inner product of two Pfaffians. Expanding them in different bases leads to two distinct versions of the one-loop KLT relations. In our letter we choose the $(n-1)!$ one-loop PT basis,
\begin{align}
\label{eq:one-loop KLT22}
  \mathcal{G}^{\text{GR}}_n
  = \langle \mathrm{Pf}|\mathrm{Pf}\rangle = \sum_{\alpha,\beta\in S_{n{-}1}}  
     \langle \mathrm{Pf}|\mathrm{PT}(1,\alpha)\rangle \,
     \big[\langle \mathrm{PT}|\mathrm{PT}\rangle^{-1}\big]_{\alpha\beta}\,
     \langle \mathrm{PT}(1,\beta)|\mathrm{Pf}\rangle = \sum_{\alpha,\beta\in S_{n{-}1}} 
     \mathcal{I}^{\text{YM}}_{n}(1,\alpha)\,
     \mathcal{K}(1,\alpha|1,\beta)\,
     \mathcal{I}^{\text{YM}}_{n}(1,\beta)\,,
\end{align}
which reproduces equation~\eqref{eq:one-loop KLT} of the main text.

Alternatively, as proposed in~\cite{He:2016mzd,He:2017spx}, one may use $n!$ tree-level $(n{+}2)$-point PT factors as the basis, leading to a representation with linear propagators,
\begin{equation}
   \mathcal{G}^{\text{GR}}_n
   = \langle \mathrm{Pf}|\mathrm{Pf}\rangle
   = \sum_{\alpha,\beta\in S_{n}}
   \langle \mathrm{Pf}|\mathrm{PT}^{\rm tree}(\alpha)\rangle \,
   \big[\langle \mathrm{PT}^{\rm tree}|\mathrm{PT}^{\rm tree}\rangle^{-1}\big]_{\alpha\beta} \,
   \langle \mathrm{PT}^{\rm tree}(\beta)|\mathrm{Pf}\rangle \,.
\end{equation}

\subsection{BCJ double copy}

For completeness, let us recall the role of the generalized Parke--Taylor (PT) factors introduced in \cite{Feng:2022wee}.  
Unlike the standard one-loop PT, these generalized factors carry explicit loop-momentum dependence through operators ${\pmb \ell}_1^{\mu}$,
\begin{align}\label{eq:pttensor}
{\pmb \ell}_1^{\mu_1}{\pmb \ell}_1^{\mu_2}\ldots {\pmb \ell}_1^{\mu_r} 
{\rm PT}(1,2,\cdots,n)
:=  \sum_{i=1}^n 
(\ell^{\mu_1} {-} k_{12\cdots i-1}^{\mu_1})(\ell^{\mu_2} {-} k_{12\cdots i-1}^{\mu_2})
\ldots (\ell^{\mu_r} {-} k_{12\cdots i-1}^{\mu_r})
{\rm PT}^{\rm tree}({+},i,i{+}1,\cdots,i{-}1,{-}) .
\end{align}
Here $k_{12\cdots i-1}=k_1+k_2+\cdots+k_{i-1}$.  
Their CHY integrals naturally yield quadratic propagators, in precise agreement with the representations used in this letter:
\begin{align}
\label{twogeneralsym}
&\frac{1}{\ell^2}\int {\rm d}\mu_n\,
{\pmb \ell}_1^{\mu_1}\ldots {\pmb \ell}_1^{\mu_r}  {\rm PT}(1,\rho(2),\cdots,\rho(n))\,
{\pmb \ell}_1^{\nu_1}\ldots {\pmb \ell}_1^{\nu_t}  {\rm PT}(1,\sigma(2),\cdots,\sigma(n))
\nonumber\\ 
\cong\;& {\rm sgn}^\rho_\sigma
\sum_{g\in T(1,\rho)\cap T(1,\sigma)}
\frac{1}{P_g} 
(\ell^{\mu_1}+k^{\mu_1}_{A(g,\rho)})\cdots(\ell^{\mu_r}+k^{\mu_r}_{A(g,\rho)})\,
(\ell^{\nu_1}+k^{\nu_1}_{A(g,\sigma)})\cdots(\ell^{\nu_t}+k^{\nu_t}_{A(g,\sigma)})\,,
\end{align}
where $\rho$ and $\sigma$ are permutations of $\{2,3,\cdots,n\}$, and the symbol $\cong$ means equivalence after loop integration.  

We introduce a uniform notation $g({\cal A}_1,\ldots,{\cal A}_m)$ for a one-loop cubic graph, where each ${\cal A}_i$ is a dangling tree attached to a corner of the polygon, and the loop momentum flows from ${\cal A}_m$ to ${\cal A}_1$.  
The product of propagators in $g$ is denoted $P_g$.  
The set $T(1,\rho)$ collects all such cubic graphs in which leg~1 sits in the first corner and the external legs follow the cyclic order $(1,\rho)$.  
The shift $A(g,\rho)$ is the momentum subset in   ${\cal A}_1$ preceding leg~1 in this ordering.  
The overall sign factor is  
$
{\rm sgn}^{\rho}_\sigma \equiv
\prod_{i=1}^{|\sigma|-1}{\rm sgn}^{\rho}_{\sigma(i),\sigma(i+1)} = {\rm sgn}_\rho^\sigma ,
$
with ${\rm sgn}^{\rho}_{i,j}=+1$ if $i$ appears before $j$ in $\rho$, and $-1$ otherwise.

Expanding a generic half-integrand in this basis,
\begin{equation}
\label{expansion}
I_n=\sum_{\rho\in S_{n-1}} {\pmb N}_{1,\rho}\,{\rm PT}(1,\rho)\,,
\end{equation}
where ${\pmb N}_{1,\rho}$ is a polynomial of ${\pmb \ell}_1$, 
the CHY construction automatically produces BCJ numerators,
\begin{equation}
\label{expansion2}
{\rm N}_{1,\rho;\ell}={\pmb N}_{1,\rho}\Big|_{{\pmb \ell}_1\to \ell}\,,
\end{equation}
satisfying Jacobi identities \cite{Feng:2022wee,Dong:2023stt}.

Conversely, once one-loop BCJ numerators are known, they can be inserted into the generalized PT expansion \eqref{expansion}.  
Starting from  \eqref{eq_Iexpand}, one finds
\begin{align}
 \mathcal{I}_{n}(\mathbb{I}) 
= \sum_{\alpha\in S_{n{-}1}} {\rm N}_{1,\alpha;\ell_1}\, \mathcal{M}(\mathbb{I}|1,\alpha) 
= \sum_{\alpha\in S_{n{-}1}} {\rm N}_{1,\alpha;\ell_1}\,  
 \langle \mathrm{PT}(\mathbb{I})|\mathrm{PT}(1,\alpha)\rangle
=
 \sum_{\alpha\in S_{n{-}1}}   
 \langle \mathrm{PT}(\mathbb{I})| {\rm N}_{1,\alpha;\ell_1}\big|_{ \ell \to {\pmb \ell}_1 } \,\mathrm{PT}(1,\alpha)\rangle .
\end{align}
This implies the equivalent expansion,
\begin{align}
\label{pfexpansion}
    |\mathrm{Pf}\rangle \cong  
     \sum_{\alpha\in S_{n{-}1}}   
| {\rm N}_{1,\alpha;\ell_1}\big|_{ \ell \to {\pmb \ell}_1 } \,\mathrm{PT}(1,\alpha)\rangle ,
\end{align}
confirming that ${\rm N}_{1,\alpha;\ell_1}$ indeed serve as BCJ numerators in the sense of \eqref{expansion}.  

Inserting this into the gravity formula gives
\begin{align}
 \mathcal{G}^{\text{GR}}_n 
 =   \langle \mathrm{Pf} |\mathrm{Pf}\rangle
 =\sum_{\alpha\in S_{n{-}1}}  \langle \mathrm{Pf}| {\rm N}_{1,\alpha;\ell_1}\big|_{ \ell \to {\pmb \ell}_1 } \,\mathrm{PT}(1,\alpha)\rangle
=\sum_{\alpha\in S_{n{-}1}} {\rm N}_{1,\alpha;\ell}\, \mathcal{I}_{n}(1,\alpha) ,
\end{align}
in agreement with \eqref{grint}.  

Expanding both Pfaffians leads to
\begin{align}
 \mathcal{G}^{\text{GR}}_n \cong  
  \sum_{\alpha,\beta\in S_{n{-}1}}    
 \langle {\rm N}_{1,\alpha;\ell_1}\big|_{ \ell \to {\pmb \ell}_1 } \mathrm{PT}(1,\alpha)\,|\, {\rm N}_{1,\beta;\ell_1}\big|_{ \ell \to {\pmb \ell}_1 } \mathrm{PT}(1,\beta)\rangle
 =\sum_{\alpha,\beta\in S_{n{-}1}} {\rm N}_{1,\alpha;\ell}\,{\rm N}_{1,\beta;\ell}\, \mathcal{M}_{n}(1,\alpha|1,\beta) ,
\end{align}
which reproduces \eqref{eq:NNM}.  
The one-loop KLT relations \eqref{eq:one-loop KLT} were already established in \eqref{eq:one-loop KLT22}, but the expansion \eqref{pfexpansion} provides additional insight: the KLT form follows from inserting a completeness relation for PT factors in the CHY representation, while the BCJ form arises from expansion in generalized PT factors.
Thus the generalized PT framework offers a direct worldsheet derivation of both the BCJ numerators and the gravity double-copy integrands constructed in the main text.
We remark that while the CHY expansion provides strong evidence for the structure of the integrands, its precise validity at the integrand level may involve subtleties beyond the scope of this appendix.

\bibliographystyle{apsrev4-1.bst}
\bibliography{Refs.bib}

\end{document}